\documentclass[secnumarabic,amssymb, nobibnotes, aps, prd]{revtex4-2}
\usepackage{graphicx}
\usepackage{amsmath}
\usepackage{subcaption}
\usepackage{xcolor}
\usepackage{multirow}

\begin{document}
	\title{RPST-Inspired Formalism for Black Holes in Flat Spacetime}
	\author{Bidyut Hazarika$^1$}
	
	\email{$rs_bidyuthazarika@dibru.ac.in$}
	\author{Prabwal Phukon$^{1,2}$}
	\email{prabwal@dibru.ac.in}	
	\affiliation{$1.$Department of Physics, Dibrugarh University, Dibrugarh, Assam,786004.\\$2.$Theoretical Physics Division, Centre for Atmospheric Studies, Dibrugarh University, Dibrugarh, Assam,786004.}
	\begin{abstract}
		In this work, we propose a formalism for the thermodynamics of flat black holes in Renyi statistics, inspired by the Restricted Phase Space Thermodynamics (RPST) framework. Our construction is motivated by the observed similarities in the thermodynamic behavior of flat black holes within the Rényi entropy framework and that of AdS black holes described by the Bekenstein entropy regime. We introduce a deformation parameter \(\lambda\) to replace the central charge \(C\). This RPST-inspired formalism incorporates \(\lambda\) and its conjugate variable, the response potential \(\zeta\), as a new pair of thermodynamic variables, analogous to the central charge \(C\) and chemical potential \(\mu\) in the AdS case. To illustrate the applicability of this formalism, we analyze two examples: the Reissner-Nordström (RN) flat black hole and the Kerr black hole.

	\end{abstract}
	\maketitle
	\section{Introduction}
	The field of black hole thermodynamics has experienced remarkable progress since its inception in the 1970s, primarily due to the pioneering contributions of Bekenstein and Hawking. Bekenstein's introduction of black hole entropy, coupled with Hawking's discovery of black hole radiation, established a profound connection between black hole physics and thermodynamics \cite{Bekenstein:1973ur,Hawking:1974rv,Hawking:1975vcx}. These seminal works demonstrated that black holes are not mere astrophysical objects but adhere to thermodynamic laws, marking a major paradigm shift in theoretical physics \cite{Bardeen:1973gs}. Building on these foundational discoveries, subsequent studies have further explored this intricate interplay \cite{Wald:1979zz,bekenstein1980black,Wald:1999vt,Carlip:2014pma,Wall:2018ydq,Candelas:1977zz,Mahapatra:2011si}.One key aspect of black hole thermodynamics is the occurrence of phase transitions, initially identified by Davies, who showed that certain points exhibit discontinuities in the heat capacity \cite{Davies:1989ey}. Another significant phase transition, the Hawking–Page transition, emerges when the black hole’s free energy changes sign, signaling a transition between distinct thermodynamic states \cite{Hawking:1982dh}. Additionally, transitions from non-extremal to extremal black hole configurations have been extensively studied across various scenarios \cite{curir_rotating_1981,Curir1981,Pavon:1988in,Pavon:1991kh,OKaburaki,Cai:1996df,Cai:1998ep,Wei:2009zzf,Bhattacharya:2019awq}. Beyond these, black hole thermodynamics has been enriched by studies drawing analogies with van der Waals systems, providing deeper insights into phase transitions and critical phenomena \cite{Kastor:2009wy,Dolan:2010ha,Dolan:2011xt,Dolan:2011jm,Dolan:2012jh,Kubiznak:2012wp,Kubiznak:2016qmn,Bhattacharya:2017nru}.\\
	
Thermodynamic entropy in black hole physics is an important concept which is related to the surface area of the event horizon. If BHs are conceptualized as three-dimensional entities, this area-based scaling would contradict the standard notion of extensive thermodynamic entropy. Therefore, the traditional Boltzmann-Gibbs statistical approach might not adequately describe black hole thermodynamics, suggesting the need for alternative frameworks. To better  understand the underlying nature of black hole entropy, several extensions of 
Boltzmann-Gibbs statistics have been discussed in literature,\cite{Cirto,Quevedo,Tsallis,Barrow,Nojiri,Reny,Sharma,Kania0}. Over the past few decades, advancements in quantum mechanics have led to modifications in the entropy expression associated with black hole horizons. These modifications primarily involve power-law and logarithmic corrections. Logarithmic entropy, such as Shannon entropy, plays a pivotal role in quantum mechanics. However, in scenarios that deviate from the asymptotic regime, where the law of large numbers is not directly applicable, alternative models like collision entropy often become more relevant.To address these inconsistencies, Rényi introduced a generalized entropy framework that harmonizes various entropy measures. This approach also allows for modifications to the black hole area law entropy.The expression for Rényi entropy is given by 
	\begin{equation}
	S=\frac{1}{\alpha} ~ ln[1+\alpha S_0]
	\end{equation}
	Where $S_0$ is the Bekenstein entropy for black holes. $\alpha$ is the Rényi
parameter. When  $\alpha$ goes to $0$,  we obtain the Bekenstein entropy again.\\

Recent studies have shown that flat black holes within the Rényi entropy framework exhibit thermodynamic behavior strikingly similar to that of  AdS black holes described by the Bekenstein entropy regime.This similarities were initially explored in \cite{Barzi1} subsequently extended to various aspects  in \cite{Barzi2,proof3,proof4,proof5,proof6,proof7,proof8,proof9}. Furthermore, it has been demonstrated in \cite{expanding} that by employing Rényi entropy to derive the Friedmann equations, the cosmological constant \(\Lambda\) can be expressed as a function of the Rényi parameter \(\alpha\). This eliminates the need to introduce \(\Lambda\) manually into the Einstein–Hilbert action, offering a natural connection between cosmology and non-extensive entropy. Inspired by these findings, we propose a potential correspondence between the thermodynamic properties of black holes in asymptotically flat spacetimes governed by Rényi statistics and those in asymptotically Anti-de-Sitter (AdS) spacetimes characterized by Gibbs-Boltzmann statistics. These parallels naturally raise the question of whether a direct correlation can be established between the cosmological constant \(\Lambda\) and the Rényi parameter \(\alpha\). In the following discussion, we attempt to explore this relationship.\\

To begin, let us consider the metric function of a Schwarzschild black hole, which represents the simplest black hole configuration, devoid of charge, rotation, or any other external parameters. The metric function in asymptotically flat spacetime is expressed as:
\begin{equation}
		ds^2=-\left(1 - \frac{2 G M}{r} \right) dt^2+\left(1 - \frac{2 G M}{r} \right)^{-1} dr^2+r^2 d\Omega
	\end{equation}
From the metric function, the mass of the black hole as a function of event horizon radius $r_+$ is calculated as
\begin{equation}
M=\frac{r_+}{2 G}
\label{masssc}
\end{equation}
	Using the expression for Renyi entropy, we can calculate the expression for event horizon radius as
	\begin{equation}
	r_+=\frac{\sqrt{G} \sqrt{e^{\alpha  S}-1}}{\sqrt{\pi } \sqrt{\alpha }}
	\label{eventhorizon}
	\end{equation}
	Substituting eq.(\ref{eventhorizon}) into eq.(\ref{masssc}),  we obtain
	\begin{equation}
	M=\frac{\sqrt{e^{\alpha  S}-1}}{2 \sqrt{\pi } \sqrt{\alpha } \sqrt{G}}
	\label{masssc1}
	\end{equation}
	If we consider $\alpha$ to be very small, we can expand eq.(\ref{masssc1}) up to first order and neglect the higher order terms. Expanding eq.(\ref{masssc1}) upto first order, we obtain
	\begin{equation}
	M=\frac{\sqrt{  S}}{2 \sqrt{\pi }  \sqrt{G}}+\frac{\alpha  ( S)^{3/2}}{8 \sqrt{\pi } \sqrt{G}}+O\left(\alpha ^{3/2}\right)
\end{equation}	 
	Next, we consider the metric for  the Schwarzchild-AdS black hole, which is given by 
	\begin{equation}
		ds^2=-\left(1 - \frac{2 G M}{r}+\frac{r^2}{l^2} \right) dt^2+\left(1 - \frac{2 G M}{r}+\frac{r^2}{l^2} \right)^{-1} dr^2+r^2 d\Omega
		\label{series}
	\end{equation}
	where $l$ is the AdS length. From the metric function, the mass of the Schwarzchild-AdS black hole as a function of entropy $S$ is obtained to be
	\begin{equation}
	M=\frac{\sqrt{S}}{2 \sqrt{\pi } \sqrt{G}} + \frac{\sqrt{G} S^{3/2}}{2 \pi ^{3/2} l^2}
	\label{massads}
	\end{equation}
	Comparing eq.(\ref{massads}) with the first two terms in eq.(\ref{series}), we obtain
	\begin{equation}
	\alpha \approx \frac{4 G}{\pi  l^2}
	\label{alpha}
	\end{equation}
	A similar relation was obtained in ref. \cite{expanding}, where authors have shown a relation between $\alpha$ and cosmological constant $\Lambda$ as follows :
	\begin{equation}
	\Lambda \approx \pm \frac{3 \alpha \pi}{G}
	\end{equation}
	 substituting, $\Lambda=-\frac{3}{l^2}$, we obtain
	\begin{equation}
	\alpha \approx \frac{G}{\pi l^2}
	\label{ref1}
	\end{equation}
	From eq.(\ref{alpha}) and eq.(\ref{ref1})  it is clear that there exist a relation between the AdS length scale $l$ and entropy parameter $\alpha$ where the parameter $\alpha$ is found to behave like $\alpha \propto \frac{1}{l^2}$. Inspired by this relation we propose a new formalism where we can study the RPST of non AdS black holes where the AdS length scale $l$ is absent.

 Restricted Phase Space Thermodynamics (RPST) was introduced for AdS black holes \cite{rp0} as a formalism that limits the phase space by concentrating on a subset of thermodynamic variables while retaining the core principles of black hole thermodynamics. This approach has been extensively applied to diverse black hole systems \cite{rp1, rp2, rp3, rp4, rp5, rp7, rp8, rp9, kerrsen, rp10, rp11,rp12,rp13} and offers significant insights into the AdS/CFT correspondence via black hole thermodynamics. In RPST, the thermodynamic variables \(P\) and \(V\) are constrained, leading to a simplified first law of the form:  
\begin{equation}
dM = TdS + \Omega dJ + \tilde{\Phi} d\tilde{Q} + \mu dC,
\end{equation}

where the AdS radius $l$ is fixed. This restriction eliminates volume work, streamlining the analysis of black hole thermodynamic processes. 

The corresponding Euler relation is adjusted to:  
\begin{equation}
M = TS + \tilde{\Phi} \tilde{Q} + \Omega J + \mu C.
\end{equation}

These equations retain the conventional structure observed in classical thermodynamics. Notably, in the RPST framework, the gravitational constant \(G\) is allowed to vary. It is crucial to distinguish between the effects of varying \(G\) and those of varying the cosmological constant \(\Lambda\). The cosmological constant, being part of the Lagrangian density, influences the gravitational field equations when modified. In contrast, \(G\) acts as a multiplicative factor in the total action, meaning that changes to \(G\) do not affect the field equations. Furthermore, variations in \(\Lambda\) modify the geometry of spacetime, whereas changes in \(G\) do not alter the spacetime structure.\\

The Restricted Phase Space Thermodynamics (RPST) framework is inherently tied to AdS black holes due to its reliance on the presence of a cosmological constant, \(\Lambda\), which is associated with the central charge \(C\) of the dual conformal field theory (CFT). However, for non-AdS black holes, where \(\Lambda\) is absent,  a RPST-like formalism for the thermodynamics of non-AdS black holes  is proposed by the authors of ref \cite{rp5},  where they varied the Newton constant and introduced a new parameter $\mu$ chemical potential. In this work, we propose a novel RPST-inspired formalism by introducing a deformation parameter \(\lambda\), replacing the central charge \(C\). There are some differences in our approach and motivations from ref\cite{rp5}. Specifically, in our study, we employ the Renyi entropy framework, which allows us to introduce an additional pair of variables, $\lambda$ and $\zeta$, into the first law of black hole thermodynamics. Notably, within the Rényi entropy formalism, the information about this extra pair is naturally embedded in the mass expression itself, eliminating the need to introduce these variables externally. In contrast to the previous approach, our manuscript seeks to establish a relation between the Rényi entropy parameter and the degrees of freedom of the system and also focuses on the cosmological constant-like behavior of the Rényi parameter.

The parameter $\lambda$, defined as the inverse of the Rényi parameter $\alpha$, plays a role analogous to $C$, encapsulating scaling properties and sensitivity to underlying statistical or thermodynamic structures. Specifically, $\lambda$ governs how interactions or correlations deviate from the traditional Boltzmann-Gibbs framework, quantifying the degree of deviation from standard extensive statistics in entropy formalisms incorporating Rényi entropy. However, while Rényi entropy follows a nonextensive entropy formalism, the RPST-inspired thermodynamic framework we employ remains extensive. The validity of the first law together with the Euler relation, ensures that the mass $M$, is a homogeneous function of the first order, confirming the extensivity of our formalism.  \\

In systems governed by Rényi entropy, $\lambda$ can be interpreted as a measure of the richness of information or effective degrees of freedom, paralleling the central charge $C$ in AdS RPST, which controls the size of the dual bulk spacetime or the density of states in the CFT. Based on these similarities, we construct the first law of black hole thermodynamics in this RPST-inspired formalism for non-AdS black holes as follows: 
\begin{equation}
dM = TdS + \Omega dJ + \tilde{\Phi} d\tilde{Q} + \zeta d\lambda,
\end{equation}  
where $\zeta = \frac{\partial M}{\partial \lambda}$ is the conjugate potential to $\lambda$. The parameter $\zeta$, referred to as the "response potential," quantifies the system's sensitivity to changes in $\lambda$, akin to the chemical potential $\mu$ in traditional thermodynamics. Physically, $\zeta$ describes how variations in the system's scaling properties, governed by $\lambda$, influence macroscopic quantities such as mass $M$ or entropy $S$. Together, $\lambda$ and $\zeta$ form a thermodynamic-like duality, with $\lambda$ controlling the structure of entropy and degrees of freedom and $\zeta$ acting as a conjugate potential that governs the system's response to changes in $\lambda$.   This duality mirrors the $C$-$\mu$ pair in AdS RPST, where $C$ measures the density of states or information content, and $\mu$ reflects the thermodynamic adjustments due to changes in state composition. Extending this analogy, $\lambda$ encapsulates deviations from standard statistical behavior and encodes the system's effective sensitivity or scaling richness, while $\zeta$ provides a measure of how these deviations manifest in the macroscopic thermodynamics, distinguishing between repulsive and attractive interactions. Thus, while the entropy structure follows the non-extensive Rényi formalism, the RPST-like thermodynamic framework constructed here remains extensive. \\

In the following sections, we present two examples to demonstrate the formulation of RPST-inspired formalism for non-AdS black holes. First, we analyze the Reissner-Nordström (RN) charged flat black hole, exploring the various processes within this framework. Next, we apply the same approach to the non-AdS Kerr rotating black hole. Finally, in the last section, we provide our concluding remarks on the implications and insights gained from this study.

	\section{Reissner-Nordström  charged black hole}
	
	The equation of motion of the RN charged black hole  in 4-dimension is 
	$$ds^2=-f(r)dt^2+\frac{1}{f(r)}dr^2+r^2(d\theta^2+sin^2\theta d\phi^2)$$
	where the metric function $f(r)$ is given by 
	\begin{equation}
	f(r)=1-\frac{2 G M}{r}+\frac{G Q^2}{r^2}
	\end{equation}
	Here $M$ and $Q$ are the mass and charge of the black hole, respectively.
	At the event horizon radius $r=r_+$, the mass is obtained as 
	\begin{equation}
		M=\frac{G Q^2+r_+^2}{2 G r_+}
		\label{rnmass}
	\end{equation}
	We rewrite the expression of the Rényi entropy as 
	The expression for the event horizon radius as a function of Renyi entropy $S$,  in the context of RN black holes, can be written as,
	\begin{equation}
	r_+=\frac{\sqrt{G} \sqrt{e^{\alpha  S}-1}}{\sqrt{\pi } \sqrt{\alpha }}
	\end{equation}
	substituting $r_+$ in the expression of mass,  the new mass is obtained as
	 \begin{equation}
	S=\lambda~ ln[1+\frac{S_0}{\lambda}]
	\label{renyi}
	\end{equation}
	Where we defined the $\lambda$  parameter as
	\begin{equation}
	\lambda=\frac{1}{\alpha} 
	\end{equation}
	and $S_0=\frac{\pi r_+^2}{G}$ is the entropy of RN black holes in GB statistics.\\
	Using eq.(\ref{renyi}), the expression for event horizon radius $r_+$ is written as
	\begin{equation}
	r_+=\frac{\sqrt{G} \sqrt{\lambda } \sqrt{e^{S/\lambda }-1}}{\sqrt{\pi }}
	\label{neweh}
	\end{equation}
	Using eq.(\ref{neweh}) in eq.(\ref{rnmass}), the mass is reformulated as 
	\begin{equation}
	M=\frac{\pi  Q^2+\lambda  \left(e^{S/\lambda }-1\right)}{2 \sqrt{\pi } \sqrt{G} \sqrt{\lambda } \sqrt{e^{S/\lambda }-1}}
	\label{unscaled}
\end{equation}	 

	Now, inspired by the relation eq.(\ref{alpha}) and eq.(\ref{ref1})  we use the rescaling constant $\kappa$ to rewrite Newton constant $G$ and rescaled electric charge $\tilde{Q}$ as
	
	\begin{equation}
	\tilde{Q} \to \frac{\kappa ~Q}{\sqrt{G}} ~~~~~~~~~\text{and}~~~~~~~G\to \frac{\kappa ^2}{ \lambda}
	\label{rescale}
	\end{equation}
	Using eq.(\ref{rescale}) in eq.(\ref{unscaled}), the mass is rewritten as 
	\begin{equation}
	M=\frac{\pi  \overset{\sim }{Q}^2+\lambda ^2 \left(e^{S/\lambda }-1\right)}{2 \sqrt{\pi } \kappa  \lambda  \sqrt{e^{S/\lambda }-1}}
	\end{equation}
	And finally the rescaled mass $\tilde{M}=M \kappa$, is written as,
	\begin{equation}
	\tilde{M}=\frac{\pi  \overset{\sim }{Q}^2+\lambda ^2 \left(e^{S/\lambda }-1\right)}{2 \sqrt{\pi }  \lambda  \sqrt{e^{S/\lambda }-1}}
	\label{finalmass}
	\end{equation}
	
If $S$, $\tilde{Q}$, and $\lambda$ are rescaled as $S \to \beta S$, $Q \to \beta Q$, and $\lambda \to \beta \lambda$, then Eq.~(\ref{finalmass}) implies  $\tilde{M} \to \beta \tilde{M}$  which proves the first order homogeneity of
$\tilde{M}$.Using the expression for mass in eq.(\ref{finalmass}), the other quantities are calculated as
\begin{equation}
T=\frac{e^{S/\lambda } \left(\lambda ^2 \left(e^{S/\lambda }-1\right)-\pi  \tilde{Q}^2\right)}{4 \sqrt{\pi } \lambda ^2 \left(e^{S/\lambda }-1\right)^{3/2}}
\label{trn}
\end{equation}
\begin{equation}
\tilde{\phi}=\frac{\sqrt{\pi } \tilde{Q}}{\lambda  \sqrt{e^{S/\lambda }-1}}
\label{phirn}
\end{equation}
\begin{equation}
\zeta=-\frac{\left(2 \lambda +e^{S/\lambda } (S-2 \lambda )\right) \left(\lambda ^2 \left(e^{S/\lambda }-1\right)-\pi  \tilde{Q}^2\right)}{4 \sqrt{\pi } \lambda ^3 \left(e^{S/\lambda }-1\right)^{3/2}}
\label{zetarn}
\end{equation}
eqs.  (\ref{trn})-(\ref{zetarn}) clearly shows the zeroth order homogeneity of these quantities. Using this mass in eq.(\ref{finalmass}), the first law of black hole thermodynamics can be written as 
\begin{equation}
d\tilde{M}=T~d S+ \tilde{\phi}~d \tilde{Q}+\zeta ~d \lambda
\end{equation}
where $\tilde{Q}$ and $\tilde{\phi}$ is the rescaled electrical charge and electric potential respectively.The new pair we have introduced $\lambda$ and $\zeta$ are the deformation parameter and response potential,l respectively.
It is also very easy to prove the Euler relation 
\begin{equation}
E=T~S+\tilde{\phi}~\tilde{Q}+ \zeta~\lambda
\end{equation}\\
\subsection{Thermodynamic Processes}
In this section, we will analyse the allowed thermodynamic processes of the black hole while fixing any two of the independent variables. Let us first consider the $T-S$ process. To obtain the critical points, one needs to solve the following pair of equations :
\begin{equation}
\frac{d T}{d S}=0,~~~~~~~~~~~~\frac{d^2 T}{d S^2}=0
\end{equation}
Solving the above equations, the critical value of $S$ and $\tilde{Q}$ is obtained as
\begin{equation}
S_C=\lambda~  ln \left(2 \left(\sqrt{3}-1\right)\right) ~~~~~~~~~~\tilde{Q}_C=\sqrt{\frac{7-4 \sqrt{3}}{\pi }} ~\lambda
\label{critical}
\end{equation}
By substituting in the expression for $T$, the critical temperature is found to be $ T_C=0.256236.$ Using this critical value, the relative temperature is written as,
\begin{equation}
t=\frac{0.550459 \left(1.4641^s \left(-0.0717968 ~ q^2+1.4641^s-1\right)\right)}{\left(1.4641^s-1\right)^{3/2}}
\end{equation}
where 
\begin{equation}
t=\frac{T}{T_C},~~~~s=\frac{S}{S_C},~~~~~q=\frac{\tilde{Q}}{\tilde{Q}_C}
\end{equation}
The rescaled Helmholtz free energy $f=F/F_C$ is obtained to be
\begin{multline}
f=\frac{\lambda}{\biggl(1.4641^s-1.\biggr)^{3/2}}  \biggl(e^{0.762484 s} (1.10092\, -0.209858 s)+q^2 \biggl(1.4641^s (0.0150671 s+0.0790423)-0.0790423\biggr)\\+1.4641^s (0.209858 s-2.20183)+1.10092\biggr)
\end{multline}
\begin{figure}[ht]
\begin{center}
\includegraphics[width=.48\textwidth]{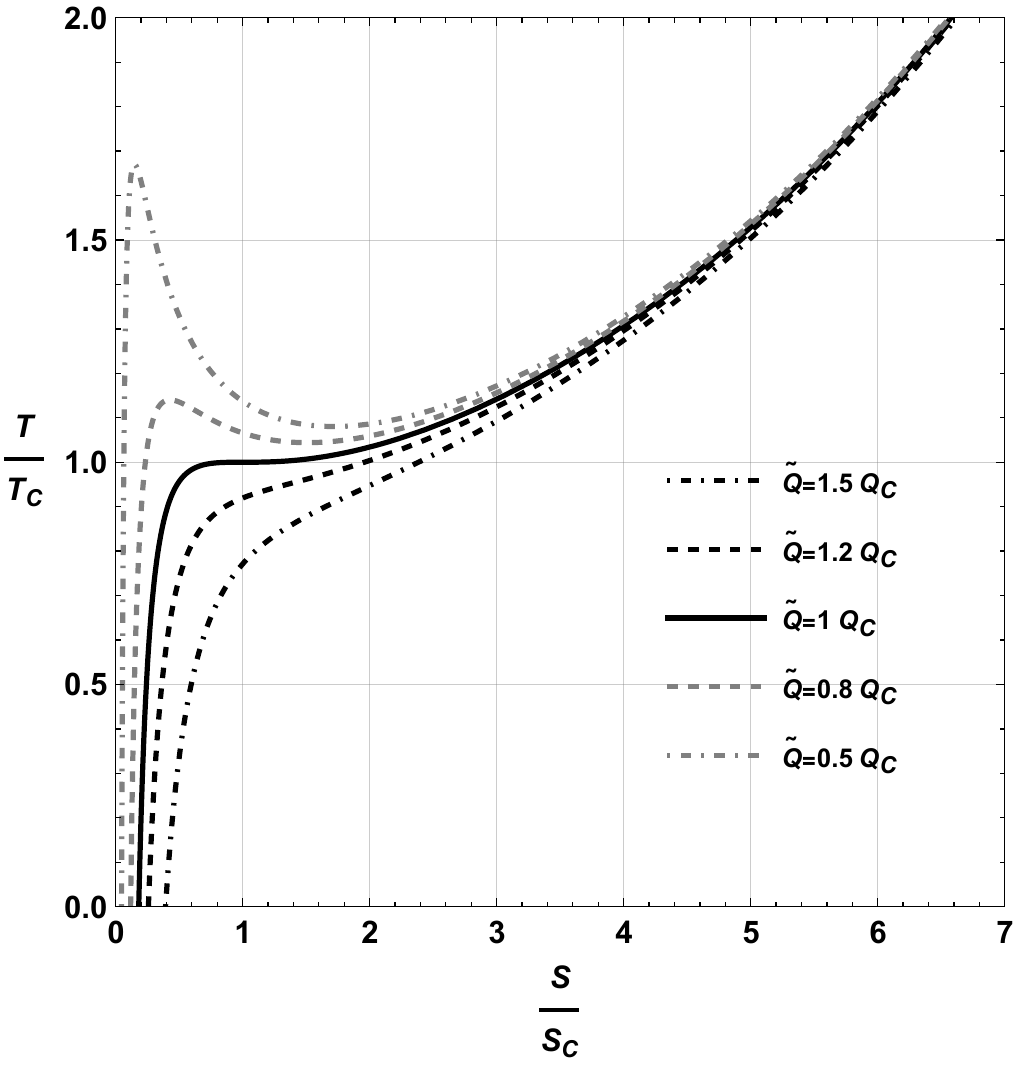}\hspace{2pt}
\includegraphics[width=.485\textwidth]{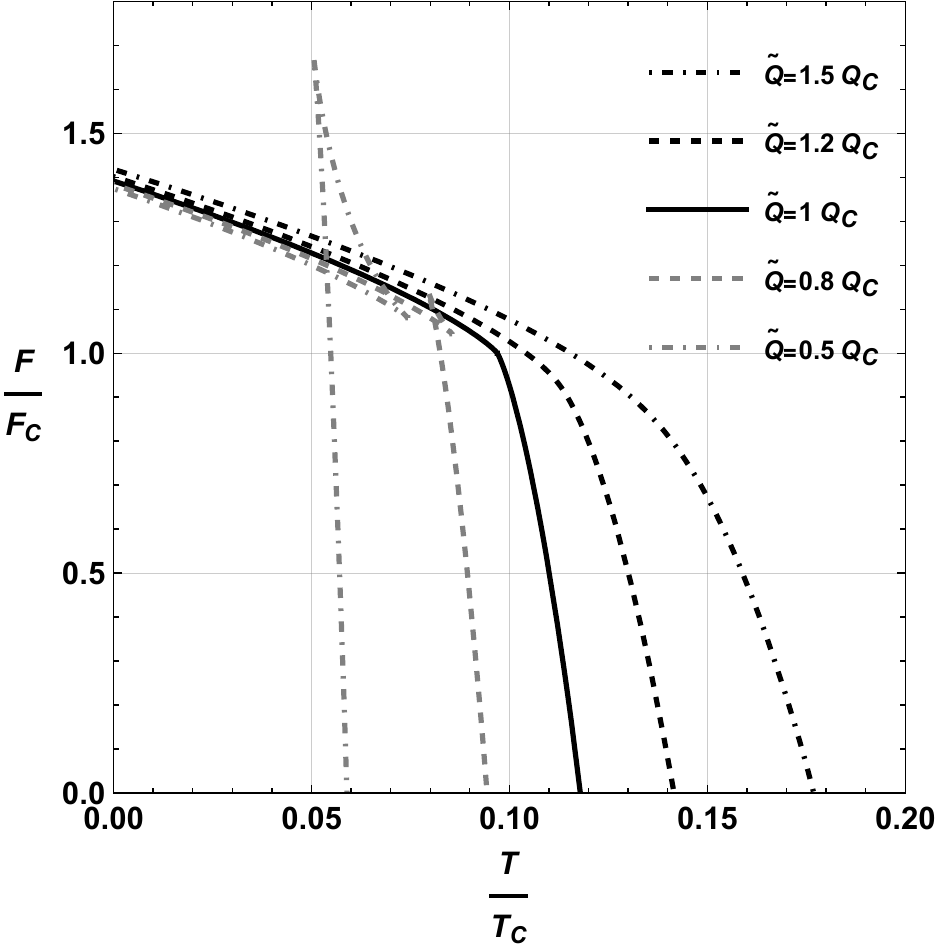}
\caption{$T-S$ and $F-T$ curves in the iso-$e$-charge processes}\label{fig1}
\end{center}
\end{figure}
FIG.  \ref{fig1} illustrates the $T-S$ and $F-T$ curves for the iso-charge processes. Below the critical values of $\tilde{Q}$, the $T-S$ curve exhibits non-monotonic behavior, and the $F-T$ curve displays a swallowtail structure. This behavior signifies a Van der Waals-like first-order phase equilibrium in the iso-e-charge processes when $0 < \tilde{Q} < \tilde{Q}_C$. At the critical point $\hat{Q} = \hat{Q}_c$, the second order phase transition is observed as represented by the black coloured solid curve in both the figures.\\

It is important to emphasize that the first-order Van der Waals-like phase transition is commonly absent in the thermodynamics of non-AdS black holes. In flat charged black holes, only Davies-type phase transitions are observed. However, when employing our RPST-like formalism, a first-order Van der Waals-like phase transition emerges, akin to what is observed in the RPST framework for charged AdS black holes.
For $\tilde{Q} = 0$, the flat charged AdS black hole reduces to the Schwarzschild black hole. Within this formalism, the Schwarzschild black hole exhibits a Davies-type phase transition. In contrast, non-AdS Schwarzschild black holes, in the absence of such a formalism, do not exhibit any phase transitions and are characterized by a single unstable black hole branch. Interestingly, under our RPST-like framework, the Schwarzschild black hole displays two branches: one stable and one unstable. This behavior closely mirrors that of the Schwarzschild-AdS black hole, providing further evidence that, even without the AdS length scale, it is possible to construct a formalism for studying the RPST of flat black holes.\\

\begin{figure}[ht]
\begin{center}
\includegraphics[width=8cm,height=7cm]{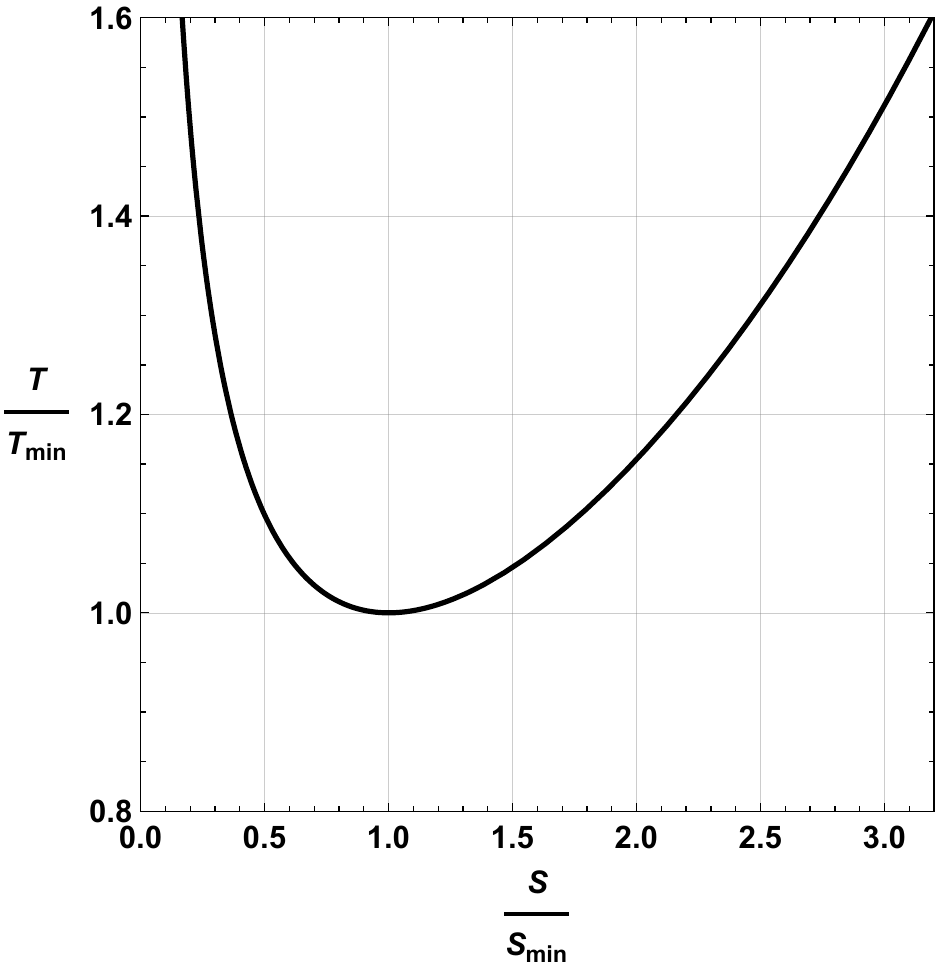}\hspace{5pt}
\includegraphics[width=8cm,height=7cm]{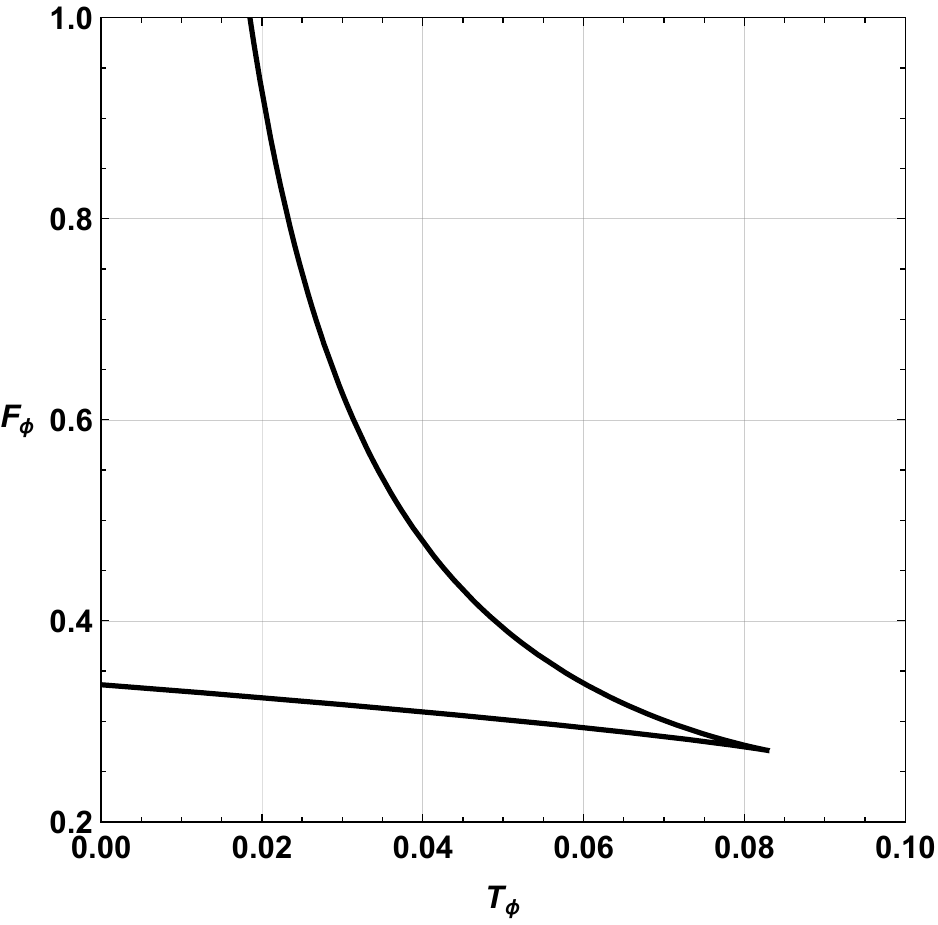}
\caption{$T-S$ and $F-T$ curves in the isovoltage processes}\label{fig2}
\end{center}
\end{figure}
Now let us consider the same $T-S$ curves in isovoltage process. To do so we use eq.(\ref{phirn}) and replace $\tilde{Q}$ in terms of $\phi.$ Considering constant value of $\lambda=1$ and $\phi=0.2$, the expression for $T$ and $F$ is plotted in FIG.\ref{fig2}. As the figures explain, there are two branches. The stability of those branches can be easily identified by plotting the heat capacity curve. It is seen that the stability of the black holes are in agreement with the principle of maximum entropy, which says a state with greater entropy ($S$) is thermodynamically favored. It is observed from the FIG. 2(a), a black hole in the state \((T, \tilde{\Phi}, \lambda)\) initially has a smaller radius, and it will transition to a state with a larger radius. At the phase transition point, the heat capacity diverges, characterizing it as a Davies-type phase transition. This type of transition is analogous to the phase transitions observed in flat charged black holes within the standard thermodynamic framework (non-extended phase space). Notably, the iso-voltage processes in our RPST formalism exhibit thermodynamic behavior consistent with that of charged black holes where the cosmological constant is absent.
\begin{figure}[ht]
\begin{center}
\includegraphics[width=8cm,height=7cm]{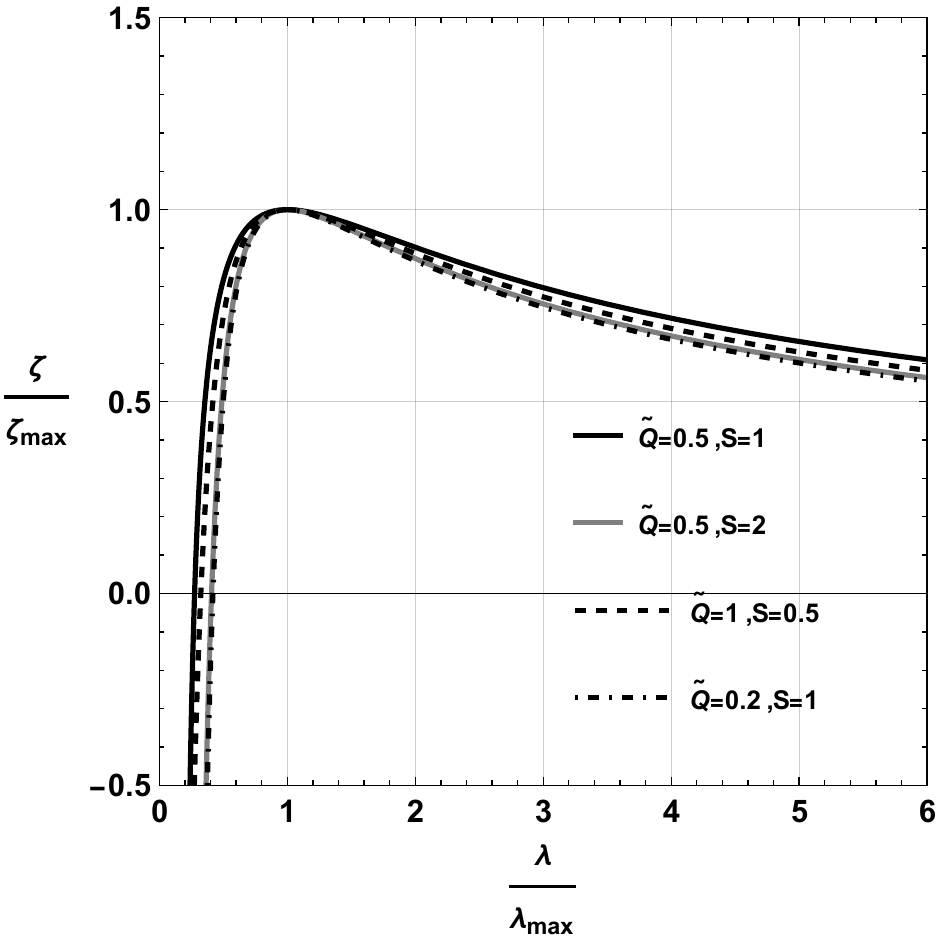}\hspace{5pt}
\includegraphics[width=8cm,height=7cm]{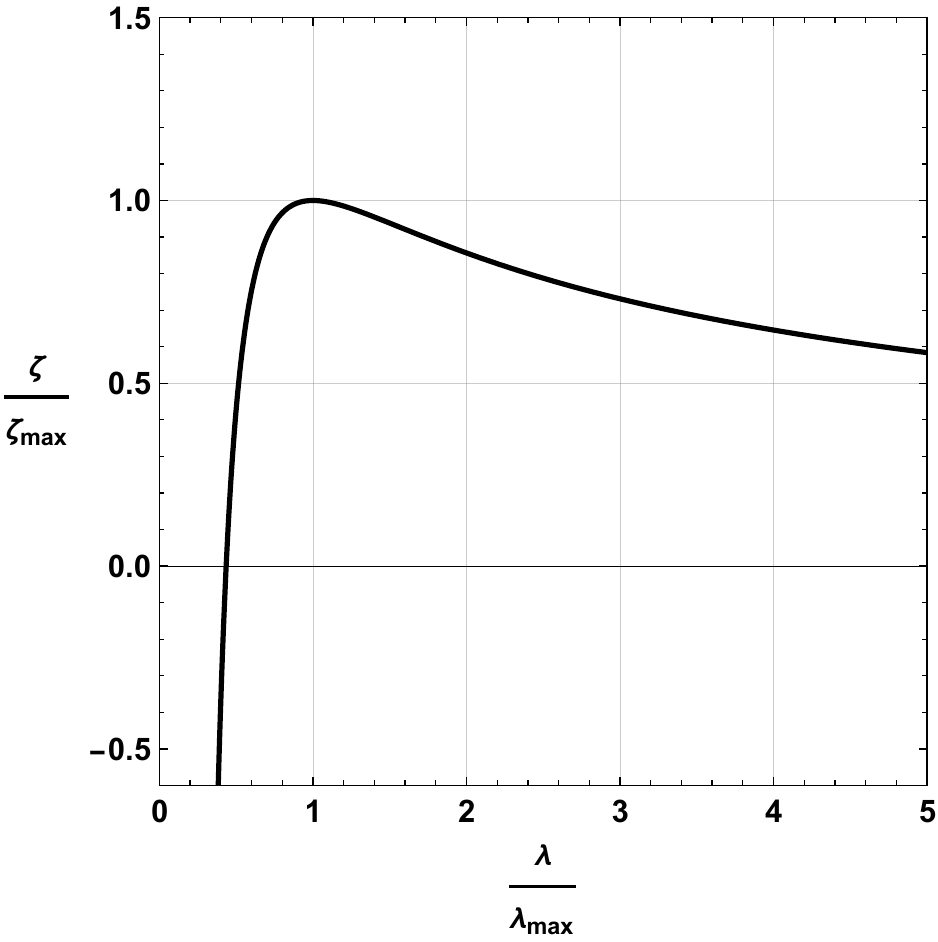}
\includegraphics[width=8cm,height=7cm]{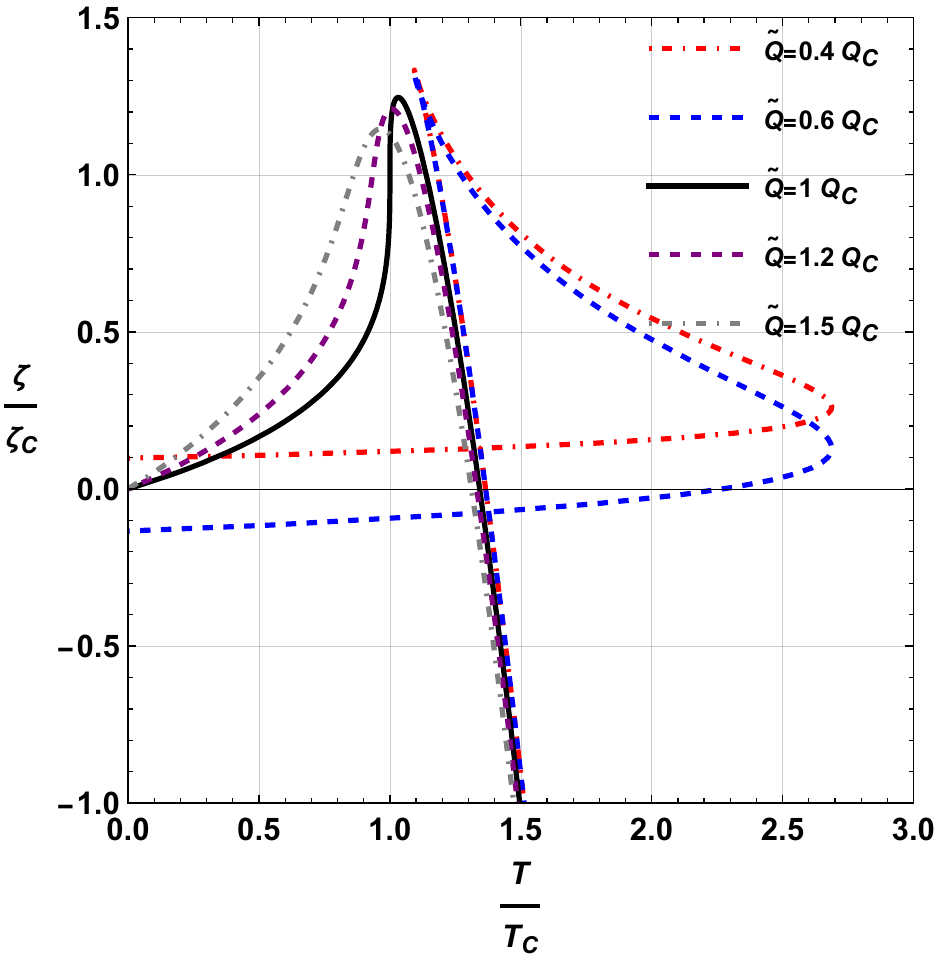}\hspace{5pt}
\includegraphics[width=8cm,height=7cm]{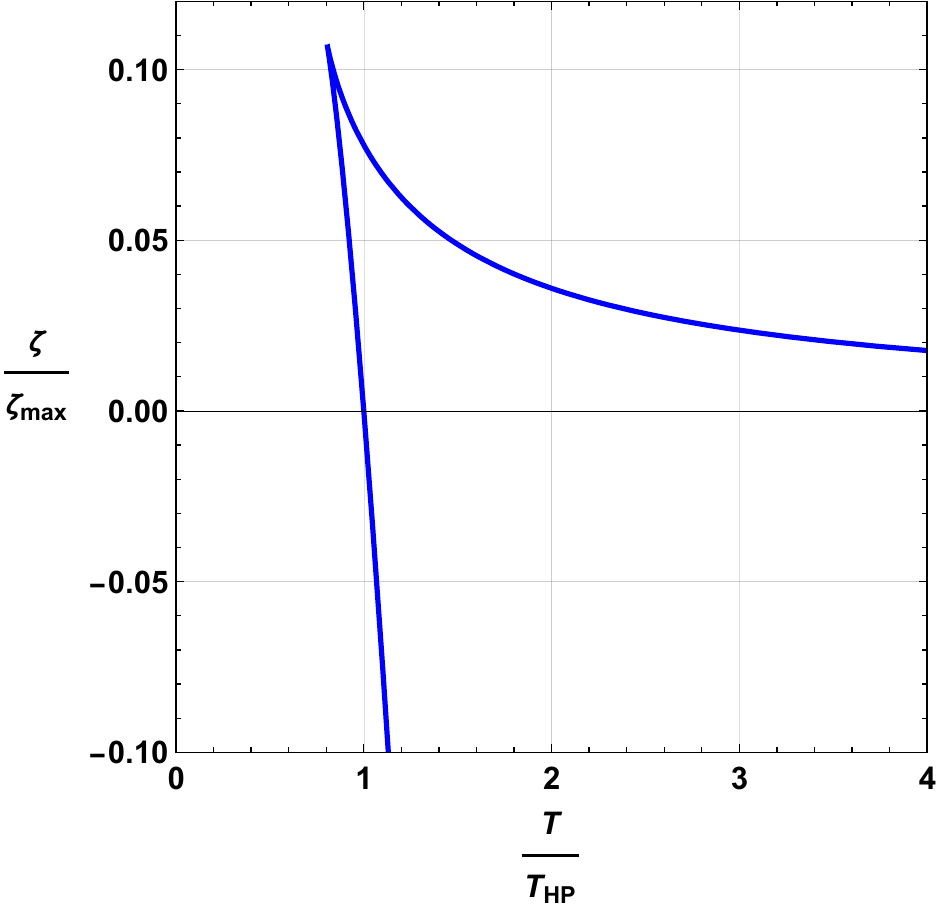}
\caption{Thermodynamic behavior of the $\zeta$-$\lambda$ and $\zeta$-$T$ processes. (a) Iso-charge $\zeta$-$\lambda$ process, showing no universal behavior. (b) Iso-voltage $\zeta$-$\lambda$ process, exhibiting universality across different $S$ and $\phi$. (c) Iso-charge $\zeta$-$T$ plot. (d) Iso-voltage $\zeta$-$T$ plot with $\phi = 0.1$ and $\lambda = 1$.}\label{fig3}
\end{center}
\end{figure}

Finally, let us analyze the $\zeta-\lambda$ process in detail. We first plot the expression for $\zeta$ as a function of $\lambda$ using Eq. (\ref{zetarn}), keeping $\tilde{Q}$ and $S$ fixed. The corresponding iso-charge process is depicted in Fig. 3(a). Unlike the $\mu-C$ process in the AdS RPST formalism, no universal behavior is observed in this case.  $\zeta = 0$ leads to the vanishing of the Gibbs free energy, $G = \zeta \lambda$. For $\zeta > 0$, the microscopic degrees of freedom of the RN black hole exhibit repulsive behavior, whereas for $\zeta < 0$, they exhibit attractive behavior.  The expression for the maximum value $\zeta_C$ can be determined by numerically solving $\frac{d \zeta}{d \lambda}=0$ for $\lambda$ while keeping $\tilde{Q}$ and $S$ constant.  Substituting $\lambda_{max}$ in the expression of $\zeta$, we finally obtain $\zeta_{max}$,  allowing us to perform a rescaling that leads to the final representation in Fig. 3(a). To further explore this relation, we also examine the $\zeta-\lambda$ process at fixed $S$ and $\phi$. This is achieved by substituting $\tilde{Q}$ as a function of $\phi$ into Eq. (\ref{zetarn}), and the resulting plot is shown in Fig. 3(b).Surprisingly,  we observe a universality in this particular scenario i.e we obtain the same $\zeta-\lambda$ plot for all set of $S-\phi$ values. Additionally, we present the $\zeta-T$ graphs for both iso-charge and iso-voltage processes in Figs. 3(c) and 3(d), respectively.  While plotting $\zeta-T$ graph for iso-voltage process we choose $\phi=0.1$ and $\lambda=1$.
In Fig. 3(c), we observe a Van der Waals-like first-order phase transition for $\tilde{Q} < Q_C$, which transitions to a second-order phase transition at $\tilde{Q} = Q_C$. Similarly, the $\zeta-T$ plots for the iso-voltage process, depicted in Fig. 3(d), reveal a Hawking-Page-like phase transition. The temperature at which $\zeta$ becomes zero is identified as the Hawking-Page temperature, $T_{HP}$. As illustrated in Fig. 3(d), for $T > T_{HP}$, the microscopic degrees of freedom exhibit repulsive behavior, whereas for $T < T_{HP}$, they exhibit attractive behavior.

	\section{Kerr black hole}
	For non AdS Kerr black hole, the mass is given by 
	\begin{equation}
	M=\frac{a^2+r_+^2}{2 G r_+}
	\label{kerrmass}
	\end{equation}
	where $r_+$ is the event horizon radius and $a=\frac{J}{M}$ is the angular momentum per unit mass.The entropy expression for this black hole is given by
	\begin{equation}
	S_0=\frac{\pi (r_{+}^2+a^2)}{G}
	\label{entropy}
	\end{equation},

	Using eq.  (\ref{renyi}),  eq.  (\ref{entropy}) in eq. (\ref{kerrmass}), the mass of the black hole in Renyi statistics can be written as
	\begin{equation}
	M=\frac{\sqrt{4 \pi ^2 J^2+\lambda ^2 \left(e^{S/\lambda }-1\right)^2}}{2 \sqrt{\pi } \sqrt{G} \sqrt{\lambda  \left(e^{S/\lambda }-1\right)}}
	\end{equation}
	By substituting $G \to \frac{\kappa^2}{\lambda}$, the rescale mass $\tilde{M}$ is finally obtained as,
	\begin{equation}
	\tilde{M}=\frac{\sqrt{4 \pi ^2 J^2+\lambda ^2 \left(e^{S/\lambda }-1\right)^2}}{2 \sqrt{\pi } \sqrt{e^{S/\lambda }-1}}
	\label{kerrfinal}
	\end{equation}
	To prove the first order homogeneity of
$M$,  we rescale $S$, $J$, and $\lambda$  as $S \to \beta S$, $J\to \beta J$, and $\lambda \to \beta \lambda$, then Eq.~(\ref{kerrfinal}) implies $\tilde{M} \to \beta \tilde{M}$.
The other quantities are calculated as :
\begin{equation}
T=\frac{e^{S/\lambda } \left(\lambda ^2 \left(e^{S/\lambda }-1\right)^2-4 \pi ^2 J^2\right)}{4 \sqrt{\pi } \lambda  \left(e^{S/\lambda }-1\right)^{3/2} \sqrt{4 \pi ^2 J^2+\lambda ^2 \left(e^{S/\lambda }-1\right)^2}}
\label{tkerr}
\end{equation}
\begin{equation}
\Omega=\frac{2 \pi ^{3/2} J}{\sqrt{e^{S/\lambda }-1} \sqrt{4 \pi ^2 J^2+\lambda ^2 \left(e^{S/\lambda }-1\right)^2}}
\end{equation}
\begin{equation}
\zeta=\frac{e^{S/\lambda } \left(4 \pi ^2 J^2 S+\lambda ^2 (6 \lambda -S)\right)-2 \lambda ^3-\lambda ^2 e^{\frac{3 S}{\lambda }} (S-2 \lambda )+2 \lambda ^2 e^{\frac{2 S}{\lambda }} (S-3 \lambda )}{4 \sqrt{\pi } \lambda ^2 \left(e^{S/\lambda }-1\right)^{3/2} \sqrt{4 \pi ^2 J^2+\lambda ^2 \left(e^{S/\lambda }-1\right)^2}}
\label{zetakerr}
\end{equation}
The zeroth order homogeneity of these quantities are quite evident.\\
Using this mass in eq.(\ref{kerrfinal}),the first law of black hole thermodynamics can be written as 
\begin{equation}
d\tilde{M}=T~d S+ \Omega dJ+\zeta ~d \lambda
\end{equation}
where $J$ and $\Omega$ is the angular momentum  and angular frequency respectively.The Euler relation can easily be proven as 
\begin{equation}
E=T~S+\Omega~J+ \zeta~\lambda
\end{equation}
We extend the analysis presented in the previous section to explore various thermodynamic processes for flat Kerr black holes. Similar to the Reissner-Nordström case, a first-order Van der Waals-like phase transition is observed in the iso-angular momentum (\(J\)) processes on the \(T-S\) plane. for critical values
\begin{equation}
S_C=0.483833 \lambda ~~~~~~~J_C=0.0193724 \lambda
\end{equation}
When the angular momentum \(J < J_C\), a Van der Waals-like phase transition occurs, and at \(J = J_C\), the transition becomes second-order. This behavior is reminiscent of the RPST framework observed for Kerr-AdS black holes. Surprisingly, for isofrequency (\(\Omega\)) processes, the \(T-S\) curve again exhibits VdW like phase transition for critical values 
\begin{equation}
S= \lambda  \log \left[\frac{1}{5} \left(3 \sqrt{6}+8\right)\right],~~~~~~~\Omega =\frac{1}{3} \sqrt{\frac{1}{2} \left(4 \sqrt{6} \pi -9 \pi \right)}
\end{equation}
Furthermore, in the \(\zeta - \lambda\) plane, we observe a Hawking-Page-like phase transition in both iso-\(J\) and iso-\(\Omega\) processes. Notably, similar to the \(\mu - C\) process in the AdS RPST formalism, the universality is absent in the \(\zeta - \lambda\) process. 
The analysis of these processes follows the same approach as detailed earlier, with equations (\ref{tkerr})–(\ref{zetakerr}) providing the necessary framework. As the procedure is straightforward and analogous to previous discussions, we omit the details here for brevity.

\section{Conclusion}

In conclusion, we have introduced a formalism for the thermodynamics of non-AdS (flat)black holes, inspired by the Restricted Phase Space Thermodynamics (RPST) framework. By drawing analogies between the thermodynamic behavior of flat black holes within the Rényi entropy framework and that of AdS black holes. In this formalism, we introduce the deformation parameter \(\lambda\) and its conjugate response potential \(\zeta\) take the place of the central charge \(C\) and chemical potential \(\mu\) pair of AdS RPST formalism. The parameter \(\zeta\), referred to as the "response potential," describes the system's sensitivity to changes in \(\lambda\), similar to the role of the chemical potential in traditional thermodynamics. Together, \(\lambda\) and \(\zeta\) form a duality where \(\lambda\) controls the entropy structure and degrees of freedom, while \(\zeta\) governs the system’s response to changes in \(\lambda\). This formalism preserves the first-order homogeneity of mass and the zeroth-order homogeneity for other thermodynamic parameters.We demonstrated the applicability of this formalism through two examples: the Reissner-Nordström (RN) flat black hole and the Kerr black hole within Einstein-Maxwell theory. \\

For flat Reissner-Nordström (RN) black holes, a first-order Van der Waals-like phase transition is observed in the iso-e-charge processes on the \(T-S\) plane when the e-charge \(\tilde{Q} < \tilde{Q}_C\). At \(\tilde{Q} = \tilde{Q}_C\), the phase transition becomes second-order. This behavior contrasts sharply with that of flat black holes, where Van der Waals-like phase transitions do not occur. The phase transitions observed in the iso-e-charge \(T-S\) plane bear a resemblance to those seen in the RPST framework for AdS charged black holes. However, for isovoltage processes, the \(T - S\) curve shows a single minimum, indicating a non-equilibrium phase transition. In the \(\zeta - \lambda\) plane, we observe a Hawking-Page-like phase transition in both iso-e-charge and isovoltage processes. Interestingly the universality is missing in the $\zeta-\lambda$ process as we have seen in the $\mu-C$ process in AdS RPST formalism. We then extend this analysis to flat Kerr black holes.\\

These analogies may be purely mathematical, or it is possible that there exists a direct relationship between the Rényi entropy and the cosmological constant. The root cause of this similarity remains an area for further exploration. If a direct connection between the cosmological constant and Rényi entropy is indeed present, as suggested by our work and a few recent literature, it could offer crucial insights into many unresolved questions in black hole physics and cosmology.\\

\section{ACKNOWLEDGMENT}
BH would like to thank DST-INSPIRE, the Ministry of Science and Technology fellowship program, Govt. of India
for awarding the DST/INSPIRE Fellowship[IF220255] for financial support.
	
	\end{document}